\documentclass{aa}  
\usepackage{graphicx}
\usepackage{txfonts}
\usepackage{xcolor}
\usepackage{float} 
\usepackage{hyperref}

\newcommand{\todo}[1]{\iffalse #1 \fi}

\newcommand{\anomalymatch}{\texttt{AnomalyMatch}}

\begin{document} 

\title{High-Redshift Gravitational Lens Discoveries in JWST NIRCam using \texttt{AnomalyMatch}}

\author{Julia Dima\inst{1}
      \and
      David O’Ryan\inst{2}\thanks{Corresponding author: david.oryan@esa.int}
      \and
        Sandor Kruk\inst{2}
      \and
      Laslo E. Ruhberg\inst{3}
      \and
      Pablo Gómez\inst{2}
      }

\institute{Polytechnic University of Timisoara, Piața Victoriei 2, Timisoara 300006, Romania \\
     \and
         European Space Agency (ESA), European Space Astronomy Centre (ESAC), Camino Bajo del Castillo s/n, 28692, Villanueva de la Cañada, Madrid
     \and
        Astronomisches Rechen-Institut (ARI), Zentrum fuer Astronomie, Universitaet Heidelberg, Moenchhofstr. 12-14, 69120 Heidelberg, Germany
         }

\date{Received ...; accepted ...}

 
  \abstract
   {Strong gravitational lenses provide a unique tool to probe cosmology and astrophysics at high redshift, offering constraints on the mass distribution of background source populations. Despite their scientific value, their rarity and subtle visual features make them challenging to identify in the wealth of data delivered by facilities such as the \textit{James Webb Space Telescope} (JWST), whose unmatched resolution and near-infrared coverage make it particularly well-suited to detecting lensing systems in this regime.}
   {We make use of the specialised open-source software \anomalymatch{}, a semi-supervised learning method to trawl the ASTRODEEP and COSMOS-Web surveys for gravitational lenses.}
   {Building on a training dataset of eleven previously identified gravitational lenses, we use \anomalymatch{} and its iterative human-in-the-loop method to train a neural network to identify gravitational lenses in \textit{JWST} Level 3 products using ESA Datalabs.}
   {In total we identify 58 unique gravitational lenses. These are graded by four experts into 16 Grade A, 16 Grade B, and 26 Grade C lenses. Of all lenses identified, 37 were previously uncatalogued. We analyse their properties such as photometric redshift measurements and spectroscopic redshift, when the latter is available. The lenses previously identified span spectroscopic redshifts to $z_{\text{spec}} \leq 1.39$ and photometric redshifts to $z_{\text{phot}} \leq 2.21$. The uncatalogued lens system with the highest redshift is at $z_\text{phot}=2.1$. }
   {Overall, we demonstrate the potential of \anomalymatch{} for large-scale searches for gravitational lenses and other rare high-redshift objects in \textit{JWST} archives.}

   \maketitle
\section{Introduction}\label{section:Introduction}
Strong gravitational lensing provides powerful insights into the distribution of mass (both baryonic and dark) in galaxies and galaxy clusters \citep[][]{2008A&A...477..397G, 2010ApJ...724..511A, 2021A&A...655A..81P, 2024ApJ...970..143T}, the evolution of galaxies \citep[][]{2015ApJ...800...18A, 2013ApJ...762...32C, 2023A&A...679A.124G}, and cosmological parameters of the Universe \citep[][]{2024MNRAS.527.5311L, 2022A&A...657A..83C}. Of particular interest are galaxy-galaxy strong gravitational lenses. These occur due to the chance alignment of two galaxies in the plane of the sky, where the gravitational potential of the foreground galaxy (the 'lens') causes the light of the background galaxy to be bent around it and be smeared into an arc. The morphology of these arcs is strongly dependent on the geometry of the lensing system, and lensing magnification additionally allows high-resolution imaging of the background source \citep{2013ApJ...772..141B, 2018ApJ...852L...7E, 2024CmPhy...7..286B, 2025MNRAS.537.1163A}. 

Therefore, identifying lensing sources at high redshifts guarantees that we can study highly magnified systems at much earlier epochs of the Universe's evolution \citep[][]{2023ApJ...952..142F, nightingale2025}. However, lensed galaxies are not the only systems of interest. Lensed transients such as supernovae can provide independent constraints on H$_{0}$ via time-delay cosmography \citep{2021MNRAS.503.1096D, 2023ApJ...959..134N, 2024SSRv..220...48B, 2026arXiv260214074A}, as can multiply-imaged quasars \citep{2007ApJ...662...62F, 2024A&A...682A.187M, 2026A&A...706L..12N}. If the lens is a galaxy cluster rather than a single galaxy, background sources are stretched into large arcs - or, in cases of near-perfect alignment, near-complete Einstein rings - enabling detailed reconstruction of the cluster's dark matter distribution \citep{2020Sci...369.1347M}.

Because gravitational lensing depends on chance alignments along the line of sight, such systems are intrinsically rare. Existing catalogues typically contain on the order of a few 10s to $\sim$1000 confirmed lensing systems \citep[e.g.][]{2022A&A...667A.141G}. For identification of further examples, we must search large observational datasets spanning either large areas of the sky or to high depths to probe a large redshift distribution.

The James Webb Space Telescope (\textit{JWST}) offers unprecedented opportunities for lens discovery due to its resolution, depth, and near-infrared coverage, which is particularly advantageous for detecting lenses at higher redshifts \citep[][]{nightingale2025}. Launched in 2021, this space observatory has built a large archive of data hosted on platforms such as the Barbara A. Mikulski Archive for Space Telescopes (MAST)\footnote{\href{https://mast.stsci.edu/search/ui/\#/jwst}{https://mast.stsci.edu/search/ui/\#/jwst}}. We focus on observations made with the Near-Infrared Camera (NIRCam) through two parallel modules each containing short ($F115W$, $F150W$ used in this work) and long ($F277W$, $F444W$ used in this work) wavelength channels. NIRCam's sensitivity to faint, high-redshift galaxies makes it well-suited for identifying strong lensing systems, while its wide set of near-infrared filters enables colour composite images in which the colour separation enhances the visual detectability of lensing features and aids automated algorithms in distinguishing the lens from the lensed source. Although \textit{JWST} is primarily a `point-and-shoot' telescope - limiting the discovery of unknown lenses in targeted observations - several wide-area surveys such as COSMOS-Web \citep[][]{casey2023}, JADES \citep[][]{2023arXiv230602465E}, and PRIMER \citep[][]{2021jwst.prop.1837D} provide ideal datasets for searching for gravitational lenses across many hundreds of thousands of sources and a wide redshift range.

Traditional methods for gravitational lens detection - including visual inspection \citep{jackson2008}, morphological/arc-finding algorithms \citep{seidel2007, bom2017}, and spectroscopic dual-redshift identification \citep{bolton2006, talbot2021, talbot2022} - are limited in their scalability and reliability. Manual inspection is hampered both by the sheer data volumes of modern surveys and by the difficulty of distinguishing true lenses from interlopers such as mergers, edge-on galaxies, and low-surface-brightness structures. Machine learning offers a compelling alternative, and since lenses are extremely rare among galaxies, their detection can be framed as an anomaly detection problem - a task extensively studied across supervised \citep{goernitz2013}, semi-supervised \citep{ruff2020}, and unsupervised \citep{daddona2020, storeyfisher2021} paradigms, with dedicated astronomical tools such as \texttt{Astronomaly} \citep[][]{2021A&C....3600481L}, \texttt{AnomalyMatch} \citep[][]{gomez2025} and \texttt{LAISS} \citep[][]{aleo2024laiss}. Large supervised models fine-tuned specifically for lens detection have also been applied across observatories, from space-based missions such as \textit{Euclid} \citep[][]{2025A&A...702A.130N, 2025NatAs...9.1116L, 2025arXiv250315324E} to ground-based surveys such as Pan-STARRS \citep[][]{2020A&A...644A.163C}. However, such supervised approaches typically require large labelled training sets, which are difficult to assemble for intrinsically rare objects such as gravitational lenses.

In this work, we leverage the newly developed open-source software \anomalymatch{} \citep{gomez2025, 2025A&A...704A.227O}, a semi-supervised active learning framework specialised in large-scale searches for efficient identification of rare objects. Through an additional human-in-the-loop learning step \citep{wu2022} and the \texttt{FixMatch} approach \citep{sohn2020}, \anomalymatch{} utilises both labelled and unlabelled data to train a neural network to present likely candidate objects of interest. Through an intuitive user interface, the user can then refine the model during training by providing additional labels on highly scored objects, iteratively improving performance. Thus, the approach facilitates rapid identification of the desired anomaly with very few initial examples.

\noindent The main contributions of this work are as follows:

\begin{itemize}
    \item We present the first application of \texttt{AnomalyMatch} to gravitational lens discovery, demonstrating its effectiveness for identifying rare morphological features in \textit{JWST} imaging.
    \item We search 614,015 individual NIRCam cutouts from the ASTRODEEP and COSMOS-Web surveys, identifying 58 lens candidates classified into Grades A, B, and C based on confidence level: 16 Grade A (high-confidence), 16 Grade B (probable), and 26 Grade C (possible) candidates.
    \item Of these, 37 candidates are previously unreported in the literature, including 9 Grade A systems. The highest-redshift newly discovered candidate lies at $z \approx 2.1$ (photometric redshift), demonstrating \textit{JWST}'s capacity to probe lensing systems to Cosmic Noon and the early Universe.
    \item We show that a search of several major surveys can be completed on the order of hours on a single GPU, highlighting the scalability of our approach to the rapidly growing volume of \textit{JWST} archival data and forthcoming wide-field surveys.
\end{itemize}

\noindent These results underscore the potential of combining semi-supervised learning frameworks with science platforms such as ESA Datalabs for systematic lens discovery in modern astronomical archives. As \textit{JWST} continues to accumulate deep, high-resolution infrared imaging, efficient and scalable detection methods will be essential to fully exploit the scientific value of gravitational lensing for cosmology and galaxy evolution studies.

This paper is laid out as follows. Section \ref{data} describes the imaging data we use, the source catalogues we use, and the source selection criteria we apply. This Section also describes the reduction process we use to create the colour cutouts that are classified into anomalous and nominal by \anomalymatch{}. We briefly describe the \anomalymatch{} method in Section \ref{methods}, as well as a brief overview of the training set used. Section \ref{results} presents the gravitational lens systems we find, including a breakdown of those already known and new as well as their redshift distribution. We discuss these results in Section \ref{discussion} and describe our limitations in detecting gravitational lenses through these surveys as well as the effectiveness of \anomalymatch{} to this specific problem. Finally, Section \ref{conclusion} concludes our work.

\section{Data}\label{data}
\noindent In this work, we use three-channel colour images with our machine learning method to classify sources into gravitational lensing and non-lensing systems. We use observations from the archive of the NIRCam instrument from \textit{JWST}. In this section, we detail how we selected sources from the archive, the observations used as well as the steps taken to create the three-band cutouts.

\subsection{Source Selection}
\noindent To identify unknown lenses, we focus on wide-area surveys conducted with \textit{JWST} rather than individually targeted observations, which are typically focused on already known systems and therefore offer limited opportunity for serendipitous discovery. Surveys, by contrast, are agnostic to observation intentions and can contain many previously unknown lensing systems across a wide range of redshifts. Furthermore, \textit{JWST}'s greater sensitivity compared to other facilities such as \textit{Euclid} means it probes to higher redshifts, making it particularly well-suited to discovering lenses towards Cosmic Noon and the early Universe.

We make extensive use of the \texttt{ASTRODEEP}-\textit{JWST} photometric catalogues \citep[][]{merlin2024}. \texttt{ASTRODEEP} itself is not a single survey in \textit{JWST}, but is a compilation of many different surveys conducted with \textit{JWST} and then re-analysed for photometric information. The \texttt{ASTRODEEP} catalogues\footnote{ASTRODEEP catalogues found here: \url{http://www.astrodeep.eu/}} include five different surveys conducted with \textit{JWST} but initially observed with the \textit{HST}. The surveys included in the ASTRODEEP photometric catalogues are: the Abell 2744 field \citep[combining several sub-fields;][]{}, the Cosmic Evolution Early Release Science Survey \citep[CEERS;][]{2023ApJ...946L..13F}, JADES, the Next Generation Deep Extragalactic Exploratory Public survey \citep[NGDEEP;][]{2024ApJ...965L...6B}, and the PRIMER-COSMOS and PRIMER-UDS survey. By combining each of these surveys, the ASTRODEEP photometric catalogues contain 531,173 sources with a total sky coverage of $\sim$0.2 square degrees. 

The advantage of using ASTRODEEP photometric catalogues over the individual surveys is that \citet{merlin2024} have conducted the source identification throughout each field using a standardised pipeline. They reduce the observations from each campaign, and then use a weighted stack of the $F356W$ and $F444W$ bands for detection. This was found to work well for further identification of deep, low surface brightness objects at high redshift. \citet{merlin2024} then applied a signal-to-noise (S/N) ratio of $\sim$2 to define a detection. While this is a low threshold, it ensures we retain completeness in our sample while \anomalymatch{} is able to handle the removal of unresolved sources downstream. The uniformity in the reduction of observations and focus on identification of faint features is perfect the identification of gravitational lenses at high redshift. Conducting cross matching between the ASTRODEEP photometric catalogue and the observations available in the archive on ESA Datalabs yields $470,732$ sources.

In this work, we also search the area of the Cosmic Evolution Survey \citep[COSMOS;][]{2007ApJS..172...38S} observed with NIRCam for gravitational lensing systems. The full COSMOS survey covers $2 \text{ deg}^2$. This area is very well explored, with publicly available spectroscopy and imaging that spans a wide wavelength range, from X-ray to radio. This smaller area observed with NIRCam, and released as COSMOS-Web \citep[][]{casey2023}, covers a contiguous area of $0.54 \text{ deg}^2$. The photometric catalogue we use contains $399,980$ sources \citep[fully described][]{2025A&A...704A.339S}. Once this is cross matched with observations available in ESA Datalabs, we identify $393,131$ sources.

\begin{figure}
    \centering
    \includegraphics[width=0.95\linewidth]{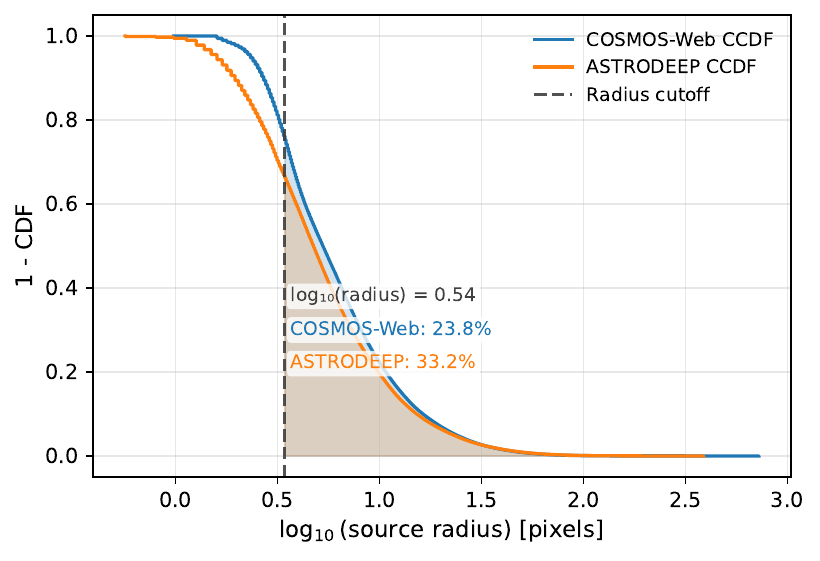}
    \caption{The CCDF of segmentation areas from the ASTRODEEP and COSMOS-Web photometric catalogues. We adopt a radial cut based on the segmentation area of each source in pixels. We retain only sources with $3.5\,r_{\mathrm{eff}} \geq 12$ pixels, equivalent to $r_{\mathrm{eff}} \geq 3.43$ pixels, or $\log(r_{\mathrm{eff}})\approx 0.54$. }
    \label{fig:ccdf}
\end{figure}

However, inspection of these catalogues revealed a large number of small, unresolved objects at high redshift. To remove such sources, we compute an effective source radius from the Source Extractor segmentation area (reported in pixels) as $r_{\mathrm{eff}}=\sqrt{\mathrm{seg\_area}/\pi}$ and define the cutout size as $3.5\,r_{\mathrm{eff}}$. We then discard sources with cutout sizes below 12 pixels, corresponding to a minimum effective radius of $\sim$3.43 pixels. For \textit{JWST}/NIRCam, this corresponds to an apparent angular radius of $\sim$0.11$^{\prime\prime}$ in the short-wavelength bands and $\sim$0.22$^{\prime\prime}$ in the long-wavelength bands. Figure \ref{fig:ccdf} shows the complementary cumulative distribution function (CCDF) of effective source radii derived from the ASTRODEEP and COSMOS-Web catalogue.

We we wish to retain the maximum number of sources from the ASTRODEEP and COSMOS-Web catalogues. While the current applied cut of a small segmentation area includes some unresolved sources, which are filtered out by our method (see Section \ref{methods}). Applying this size cutoff leaves 314,449 sources (removing 33.2\%) from ASTRODEEP and 299,566 sources (removing 23.8\%) from COSMOS-Web. This results in of 614,015 total sources across the 0.74 square degrees of the ASTRODEEP and COSMOS-Web surveys. Table \ref{tab:source-split} summarises the distribution of sources across the different surveys making up the ASTRODEEP catalogue and COSMOS-Web.

\begin{table}[]
    \centering
    \caption{Number of sources we search for gravitational lenses after applying our selection criteria.}
    \label{tab:source-split}
    \begin{tabular}{cc}
    \hline
    \hline
        Field & Sources \\
    \hline
        Abell 2744 & 33,460 \\
        CEERS & 27,191 \\
        JADES & 90,162 \\
        NGDEEP & 9,874 \\
        PRIMER & 153,762\\
        COSMOS-Web & 299,566 \\
    \hline
        Total & 614,015 \\
    \hline
    \end{tabular}
\end{table}

\subsection{Available Data on ESA Datalabs}
\noindent To create three-channel RGB images of each of our sources, we use the platform ESA Datalabs \citep{navarro2024} to directly access the \textit{JWST} archives. ESA Datalabs is a science platform that enables direct analysis of large quantities of data from the reduction pipelines. For this, the platform provides preconfigured Python environments (specifically, JupyterLab), which have data volumes mounted containing all archival, public \textit{JWST} data. We crossmatch between the ASTRODEEP source catalogues and all the calibration Level 3 data.

The calibration Level 3 data in the \textit{JWST} archives are science-ready mosaics output by the \textit{JWST} reduction pipeline. We compile the observational footprints of all available mosaics in the $F115W$, $F150W$, $F277W$, and $F444W$ bands for the ASTRODEEP and COSMOS-Web fields into a single catalogue, which we use to cross-match against our 614,015 sources using \texttt{Astropy}. For each source, we identify which mosaic footprint it falls within and record the corresponding file path, enabling direct cutout extraction across all four filters. The resulting single-band cutouts are then combined into three-channel RGB images, as described in Section \ref{subs:cutout_creation}. Note that the short-wavelength channels ($F115W$, $F150W$) have approximately half the pixel scale of the long-wavelength channels and we therefore upsample the long-wavelength cutouts to a common resolution prior to combination.

\subsection{Training Set}
To train a model, we require a training set of example lenses in JWST. For this, we turn to the COSMOS-Web Lens Survey \citep[COWLS;][]{nightingale2025}. COWLS was a gravitational lens search conducted using visual inspection by multiple experts across 42,660 sources, identifying 419 candidate lenses. For our training set, we select their sample of `spectacular' lenses \citep[][]{2025MNRAS.544L...8M}. This is a subsample of 17 lenses they identify, and on initial inspection, over half of the 32 experts classified them as lenses. Cross matching these 17 spectacular lenses with the Level 3 \textit{JWST} archive available on ESA Datalabs provides us 11 of these lenses to use as training for \anomalymatch{}. However, the semi-supervised methodology of \anomalymatch{} means that this small training set is sufficient to train the model effectively.

\section{Methods}\label{methods}
Here, we describe the steps taken to prepare our data and conduct our search for gravitational lenses. We detail the creation of three-channel RGB colour cutouts from the four \textit{JWST} NIRCam filters ($F115W$, $F150W$, $F277W$, and $F444W$) derived from the Level 3 archive data. We provide an overview of the \anomalymatch{} semi-supervised framework and describe how it is trained and applied to search the ASTRODEEP and COSMOS-Web surveys for gravitational lens candidates. Finally, we describe the expert grading system used to assess confidence in each identified candidate.

\subsection{Cutout Creation}\label{subs:cutout_creation}
\noindent Having matched our source catalogues to the FITS files, we create three-channel imaging cutouts for classification by \anomalymatch{}. We centre each cutout on the source coordinates from the \texttt{ASTRODEEP} catalogue, and estimate the cutout size from the \texttt{ISOAREA\_SE} column (the segmentation area output by SourceExtractor) assuming circular morphology to derive an effective source radius $r$. Each cutout is created with a size $3.5r\times3.5r$ to capture the relevant visual vicinity. Due to resolution differences between the short ($F115W$, $F150W$) and long wavelength ($F277W$, $F444W$) channels, we upsample the resolution of the long wavelength cutouts by a factor of 2. Each band cutout is then rescaled to $224\times224$ pixels prior to classification.


Upon the creation of each band cutout, we combine them to a three-channel image (in JPEG format) to be used in our machine learning method. First, we scale the dynamic range of each band cutout individually using a mid-tones transfer function (MTF). This method has been used successfully in creating colour cutouts of sources in the \textit{Euclid} survey \citep[][]{euclidq1}. MTF differs from other methods of normalising the images as it balances the dynamic range of the image between the low surface brightness features of an object and the high surface brightness central areas. This is done by boosting the visibility of low surface brightness, while not saturating bright pixels. In our case, looking for potentially low surface brightness gravitational lenses, which will be close to the bright cores of a galaxy, MTF is perfect for our image creation purposes.

We initially calculate the mid-point parameter $m$ such that the mean of the transformed image matches a target mean. This is done by applying

\begin{equation}\label{eqn:mtf-calc}
m \;=\; \frac{\bar{x}\,(1-\alpha)}{\bar{x}\,(1-2\alpha)+\alpha}
\end{equation}

\noindent where $m$ is the midpoint parameter for the normalised cutout (i.e. the input value that maps to 0.5), $\bar{x}$ is the mean of the original cutout. $\alpha$ is the target normalised median of the cutout -- i.e. the normalised median of the image we want after we have applied the MTF to each cutout. Through experimentation with different target medians, we empirically find $\alpha\in\{0.27,0.25,0.23,0.25\}$ for $F115W$, $F150W$, $F277W$, $F444W$, respectively. We then apply the MTF per pixel using

\begin{equation}
    MTF(x;m) \;= \dfrac{(m-1)\,x}{(2m-1)\,x - m}
\end{equation}

\noindent where $MTF(x;m)$ is the normalised per pixel value. Example code to conduct such an MTF transform is publicly available on GitHub\footnote{\url{https://github.com/mwalmsley/bulk-euclid-cutouts}}.

Secondly, we need to combine our normalised four bands into three channel JPEGs. Rather than simply combine two filters into one channel, we follow an approximation of the cutout combination process described briefly in \citet{2025ApJ...987...74G} and will be fully detailed in Galaxy Zoo et al. (in prep.). In this process, each channel is a weighted combination of all four filters. We use the following weights to create our RGB images

\begin{equation} \label{eqn:weighting}
    \begin{split}
        B &= 0.3\, F_{F277W} + 0.7\, F_{F444W} \\
        G &= 0.5\, F_{F150W} + 0.5\, F_{F277W} \\
        R &= 0.7\, F_{F115W} + 0.3\, F_{F150W} 
    \end{split}
\end{equation}

\noindent where F$_{\text{\textit{JWST} filter}}$ is the normalised flux in the relevant \textit{JWST} filter. These weights were determined empirically by experimenting with different combinations to achieve the best balance between the bright galaxy centre, low surface brightness lensing features, and the sky background.

\begin{figure*}
    \centering
    \includegraphics[width=0.95\textwidth]{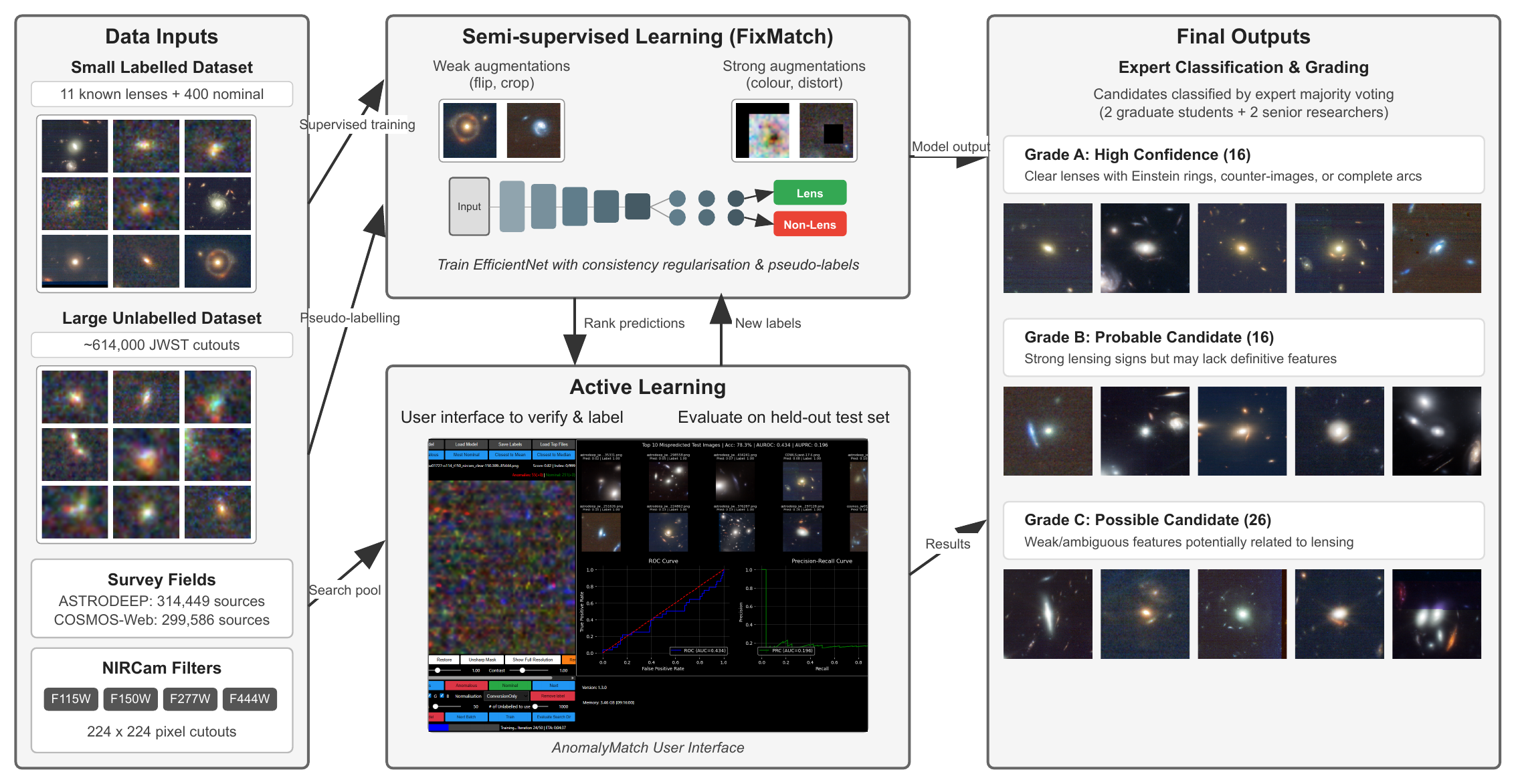}
    \caption{The AnomalyMatch workflow, starting with constructing and pre-processing dataset followed by the semi-supervised training loop and label proposals. The expert validation step involves confirming previously identified lenses based on current research and grading newly discovered lens candidates, as detailed in Section~\ref{subs:lens_grading}.}
    \label{fig:workflow}
\end{figure*}

\subsection{\anomalymatch{}}
To conduct our search, we use the \anomalymatch{} semi-supervised machine learning method \citep{gomez2025}.  \anomalymatch{} is a neural network-based approach optimised for large-scale search, training on heavily imbalanced datasets - as typical for anomaly detection tasks - and iterative, user-friendly searches. The \texttt{FixMatch} method, refined in \texttt{MSMatch} \citep{2021arXiv210310368G}, that it relies on trains a neural network using a supervised cross-entropy loss function for the labelled data. This is combined with an unsupervised loss that uses consistency regularisation on pseudo-labels of differently augmented unlabelled images. It has shown success ranging from standard machine learning benchmarks \citep{sohn2020, gomez2025} to Earth observation \citep[][]{2021arXiv210310368G} and astronomy \citep{2025A&A...704A.227O}. The key benefit to this approach is that as a result of this combined loss function leveraging both labelled and unlabelled data, only a small training set (on the order of $\sim$10 lenses) is required for excellent performance. 

To further take advantage of this approach, an active learning loop has been introduced to the training of \anomalymatch{}. Once the model is trained it will display, in a guided user interface, the top scoring images identified in the unlabelled data used for training. This allows the user to assess whether the model is converging correctly - if genuine anomalies are ranked highly - or diverging, if nominal sources dominate the top scores. The user can then add these sources to the labelled training data, allowing the user unique control over direct guiding \anomalymatch{} to identify the objects they are specifically searching for.

Upon training and then searching through an entire dataset, a score is given to each image between 0 (nominal) and 1 (anomalous). It is important to note that these are not measures of probability (though they can be calibrated as such) but are scores dependent on the internal model parameters. Therefore, when searching for anomalies in our dataset, we take the top-ranked 1,000 sources and visually inspect them for gravitational lensing systems.

\anomalymatch{} is directly integrated in ESA Datalabs and is also available open-source online\footnote{\href{https://github.com/esa/AnomalyMatch}{https://github.com/esa/AnomalyMatch}}. Figure \ref{fig:workflow} shows an overview of the \anomalymatch{} workflow. Even though \anomalymatch{} does not specialise in specific anomalies, it is particularly well-suited to lens discovery as it effectively picks up even faint visual features, which already enabled the discovery of 140 lenses in \textit{Hubble} Legacy Archive data \citep[of which 86 were uncatalogued][]{2025A&A...704A.227O}. 

\subsection{Training \anomalymatch{} and Active Learning}
We employed the \anomalymatch{} software to systematically identify likely gravitational lens candidates in \textit{JWST} imaging by combining large-scale cutout creation, semi-supervised anomaly detection, and expert labels. Starting from the ASTRODEEP and COSMOS-Web source catalogues, we generated $224 \times 224$ pixel cutouts rescaled to the NIRCam short-band scale, as specified in Section \ref{subs:cutout_creation}. 

\begin{figure}
    \centering
    \includegraphics[width=0.95\columnwidth]{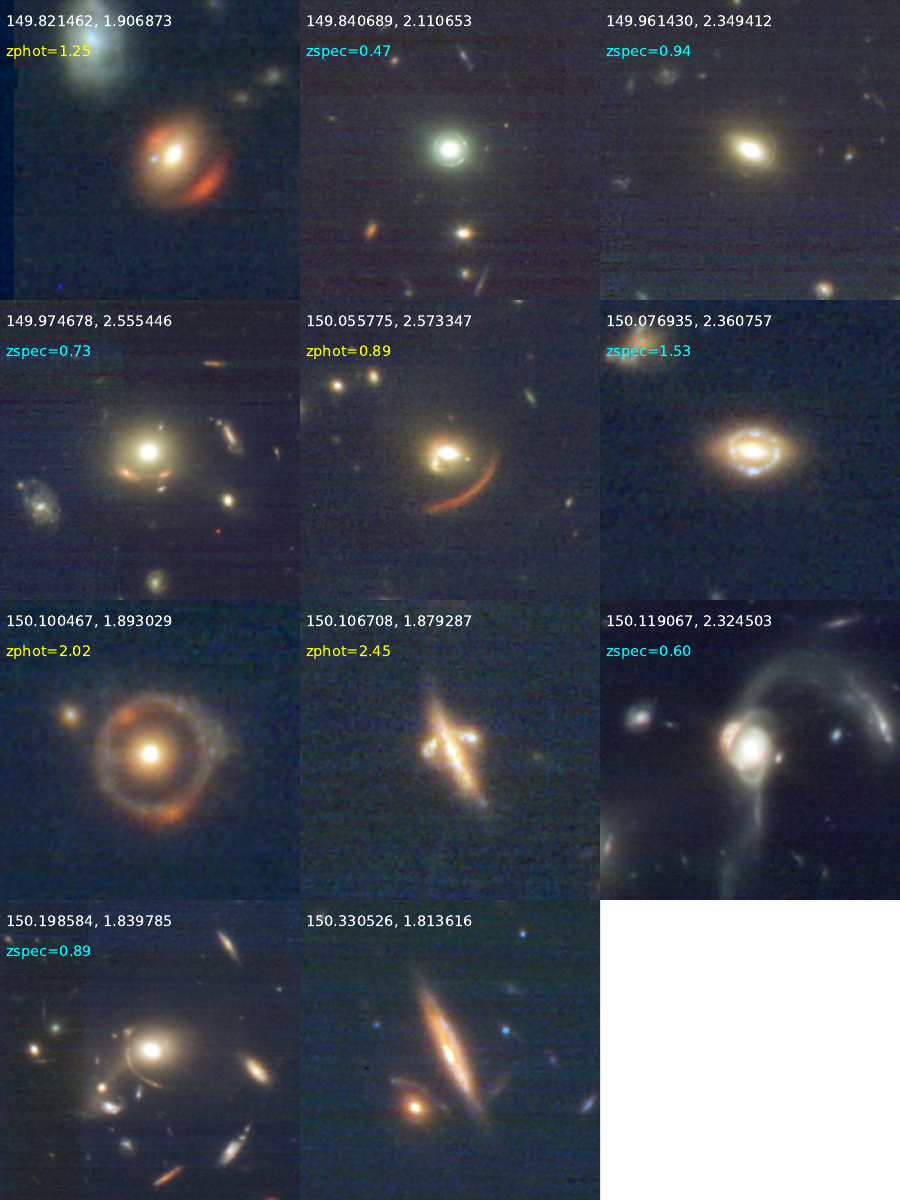}
    \caption{Initial lens dataset utilized in the \anomalymatch{} semi-supervised loop, containing 11 lenses from \citet{2025MNRAS.544L...8M} cross matched in Level 3 \textit{JWST} data.}
    \label{fig:labels_COWLS_initial}
\end{figure}

Figure \ref{fig:labels_COWLS_initial} shows the lenses used in our training set with the cutouts created using our reduction pipeline. We also visually inspect and provide 400 nominal images - i.e. non-lens sources - from ASTRODEEP as part of our training set. Therefore, our initial training of \anomalymatch{} consists of 411 sources.

We retrain \anomalymatch{} over three training cycles, each running $250$ training iterations and using $20,000$ unlabelled cutouts for unsupervised learning per cycle. Performing an initial search on ASTRODEEP, we identified a first batch of lens candidates from the $1,000$ likeliest anomalies classified by the model, which were added to our labelled dataset. We then performed a second iteration of training a model and searching, including the COSMOS-Web cutouts, following the same training approach. We repeated this multiple times until the model did not identify any new lenses.

\subsection{Lens grading system}\label{subs:lens_grading}
Once we had identified a set of lenses in the \textit{JWST} archives, we aimed to quantify our confidence in each lens. We adopted a classification scheme based on previous works conducting lens searches \citep[e.g.][]{2022A&A...667A.141G, 2025arXiv250315324E}. This involved three distinct grades from A to C described as:  

\begin{itemize}
    \item Grade A - clear arc with distinct colour from central source, typically with counter-image;  
    \item Grade B - plausible candidates lacking counter-images or showing unusual arc geometry;  
    \item Grade C - weak candidates with unusual features, potentially non-lenses.  
\end{itemize}

This classification was carried out on Zooniverse\footnote{https://www.zooniverse.org/projects/ori-j/galaxy-lens-class-voting} by four experts in searching for gravitational lenses. A majority vote was used to give the final grade of the lens. In the event of a tie, the lower grade was assigned to the image.

\section{Results}\label{results}
\noindent \anomalymatch{} was trained with our initial training set. We conduct three rounds of active learning on this model, using our labelled training sample (11 confirmed lenses and 400 non-lenses) and $20,000$ unlabelled images. After each training session, we took the top-scored 1,000 sources and visually inspected them for gravitational lenses. Any gravitational lenses found as part of this active learning are part of the results here, as they were not initially known before training. We repeat this process three times. After the final round of training was conducted with this sample and 20,000 unlabelled images before the model was applied to all of our selected sources in ASTRODEEP and COSMOS-Web.

We visually inspected the top 1,000 ranked sources to search for lenses. In total, we identified 58 lenses in the two surveys that were not included in the training set. Table \ref{tab:lenses} provides the source ID, right ascension, declination, redshift, redshift measurement method, a lens grade, and which catalogue the lens was discovered in. The source ID has the format of (\texttt{JWST\_SURVEY})\_(\texttt{ASTRODEEP\_SOURCEID}) for the specific survey where a lens was found. However, some of the gravitational lenses identified were background sources in the larger cutout of a foreground source. Therefore, these sources did not have an official source ID. We then leave these blank as '-' and quote their right ascension and declination.

\begin{table*}
\caption{Gravitational lens systems identified in this work. Systems are grouped by grade (A–C), with newly reported lenses listed first within each grade and ordered by decreasing $z_\mathrm{lens}$. Gravitational lens systems in the background of cutouts have no source ID and marked as '-'. We also state the source catalogue each was found in, with `COWLS' designating that the lens was previously identified in COWLS.}
\label{tab:lenses}
\centering
\begin{tabular}{lrrllrrl}
\hline
\hline
ID & RA & DEC & New & $z_{\text{lens}}$ & $z_{\text{type}}$ & Grade & Catalogue \\
\hline
-- & 149.794040 & 2.008240 & False & 1.414 & zphot & A & COSMOS \\
CEERS\_56193 & 214.888879 & 52.883553 & False & 1.060 & zspec & A & ASTRODEEP \\
JADES-GS\_14262 & 53.100356 & -27.870796 & False & 1.013 & zspec & A & ASTRODEEP \\
PRIMER-COSMOS\_8436 & 150.092521 & 2.202988 & False & 0.674 & zspec & A & ASTRODEEP \\
PRIMER-UDS\_50243 & 34.275995 & -5.221575 & False & 0.628 & zspec & A & ASTRODEEP \\
PRIMER-UDS\_753 & 34.340047 & -5.316749 & False & 0.625 & zphot & A & ASTRODEEP \\
ABELL2744\_15500 & 3.554197 & -30.376287 & False & 0.309 & zspec & A & ASTRODEEP \\
ABELL2744\_25433 & 3.596040 & -30.352626 & False & 0.254 & zspec & A & ASTRODEEP \\
ABELL2744\_21147 & 3.578037 & -30.359454 & False & 0.200 & zphot & A & ASTRODEEP \\
COSJ095924+020807 & 149.851139 & 2.135455 & True & 1.136 & zspec & A & COWLS \\
PRIMER-UDS\_49028 & 34.404784 & -5.224862 & True & 0.646 & zspec & A & ASTRODEEP \\
PRIMER-COSMOS\_26353 & 150.183582 & 2.252408 & True & 0.581 & zspec & A & COWLS \\
-- & 149.874717 & 2.231156 & True & 0.422 & zphot & A & COSMOS \\
PRIMER-COSMOS\_46427 & 150.100094 & 2.297128 & True & 0.362 & zspec & A & COWLS \\
COSJ095912+020750 & 149.801604 & 2.130819 & True & 0.330 & zphot & A & COWLS \\
ABELL2744\_13454 & 3.609545 & -30.3820099 & True & 0.250 & zphot & A & ASTRODEEP \\
\hline
PRIMER-UDS\_44433 & 34.318519 & -5.237388 & False & 1.390 & zspec & B & ASTRODEEP \\
-- & 3.495370 & -30.361276 & False & 1.325 & zphot & B & ASTRODEEP \\
JADES-GS\_36896 & 53.097066 & -27.823978 & False & 1.158 & zspec & B & ASTRODEEP \\
ABELL2744\_4459 & 3.584620 & -30.421733 & False & 0.775 & zphot & B & ASTRODEEP \\
-- & 3.536551 & -30.361421 & False & 0.700 & zphot & B & ASTRODEEP \\
PRIMER-UDS\_72509 & 34.393328 & -5.182571 & False & 0.496 & zspec & B & ASTRODEEP \\
ABELL2744\_22525 & 3.499966 & -30.359605 & False & 0.317 & zspec & B & ASTRODEEP \\
ABELL2744\_19792 & 3.620714 & -30.369540 & False & 0.275 & zphot & B & ASTRODEEP \\
ABELL2744\_31112 & 3.557337 & -30.340738 & False & 0.250 & zphot & B & ASTRODEEP \\
JADES-GN\_27128 & 189.205931 & 62.229689 & False & 0.089 & zspec & B & ASTRODEEP \\
JADES-GN\_19566 & 189.236016 & 62.205592 & True & 0.954 & zspec & B & ASTRODEEP \\
-- & 3.577022 & -30.379466 & True & 0.498 & zspec & B & ASTRODEEP \\
COSJ095940+023253 & 149.918706 & 2.548231 & True & 0.472 & zphot & B & COWLS \\
ABELL2744\_11690 & 3.588142 & -30.395060 & True & 0.300 & zspec & B & ASTRODEEP \\
-- & 150.304627 & 2.055907 & True & 0.295 & zspec & B & COWLS \\
COSJ100011+021301 & 150.049692 & 2.217150 & True & 0.270 & zphot & B & COWLS \\
\hline
-- & 53.119966 & -27.842394 & False & 2.100 & zphot & C & ASTRODEEP \\
PRIMER-UDS\_83561 & 34.355060 & -5.166309 & False & 1.575 & zphot & C & ASTRODEEP \\
-- & 214.988924 & 52.990572 & False & 1.243 & zspec & C & ASTRODEEP \\
JADES-GN\_20196 & 189.217004 & 62.207309 & False & 1.224 & zspec & C & ASTRODEEP \\
JADES-GS\_8833 & 53.072869 & -27.880060 & False & 1.097 & zspec & C & ASTRODEEP \\
-- & 150.332709 & 1.868228 & False & 1.006 & zphot & C & COSMOS \\
JADES-GN\_2459 & 189.218924 & 62.155466 & False & 0.954 & zspec & C & ASTRODEEP \\
PRIMER-COSMOS\_38523 & 150.074389 & 2.280903 & False & 0.939 & zspec & C & ASTRODEEP \\
CEERS\_71105 & 214.842869 & 52.859302 & False & 0.914 & zspec & C & ASTRODEEP \\
JADES-GS\_1531 & 53.042634 & -27.904544 & False & 0.765 & zspec & C & ASTRODEEP \\
ABELL2744\_2236 & 3.641470 & -30.433118 & False & 0.750 & zphot & C & ASTRODEEP \\
-- & 150.041799 & 2.069472 & False & 0.667 & zphot & C & COSMOS \\
-- & 150.146363 & 2.061356 & False & 0.653 & zphot & C & COSMOS \\
-- & 150.083213 & 2.150062 & False & 0.411 & zphot & C & COSMOS \\
-- & 149.854701 & 2.296876 & False & 0.345 & zphot & C & COSMOS \\
ABELL2744\_23664 & 3.532148 & -30.360380 & False & 0.315 & zspec & C & ASTRODEEP \\
ABELL2744\_24307 & 3.579403 & -30.358357 & False & 0.305 & zspec & C & ASTRODEEP \\
-- & 3.624310 & -30.369530 & False & 0.295 & zspec & C & ASTRODEEP \\
-- & 150.330832 & 1.813262 & True & 2.208 & zphot & C & COWLS \\
COSJ095914+021219 & 149.811407 & 2.205425 & True & 1.053 & zspec & C & COWLS \\
JADES-GS\_32872 & 53.054271 & -27.829564 & True & 0.660 & zspec & C & ASTRODEEP \\
COSJ100004+015307 & 150.020617 & 1.885335 & True & 0.338 & zspec & C & COWLS \\
ABELL2744\_10084 & 3.574892 & -30.398356 & True & 0.318 & zspec & C & ASTRODEEP \\
ABELL2744\_13465 & 3.595497 & -30.388669 & True & 0.303 & zspec & C & ASTRODEEP \\
ABELL2744\_13110 & 3.585307 & -30.387526 & True & 0.301 & zspec & C & ASTRODEEP \\
ABELL2744\_1148 & 3.583048 & -30.433519 & True & 0.293 & zspec & C & ASTRODEEP \\
\hline
\end{tabular}
\end{table*}

\begin{figure*}
    \centering
    \includegraphics[width=0.95\linewidth]{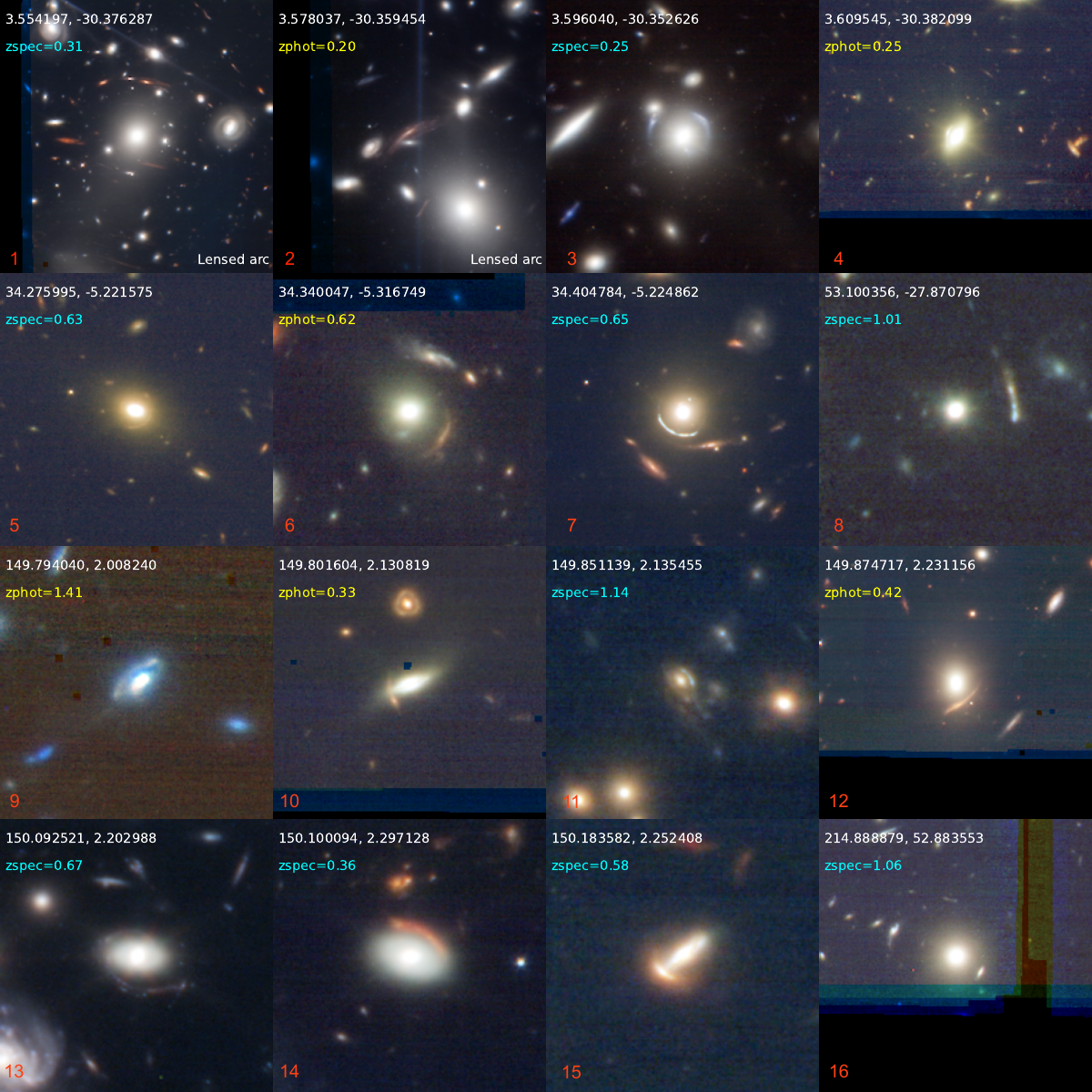}
    \caption{Mosaic of lenses given a Grade A by expert classifiers. These represent the clearest lenses, where a clear arc of a differing colour is visible with a secondary image of the arc often present. Inset to each cutout is the right ascension and declination, including a measure of the redshift with its measurement method specified (spectroscopic vs photometric). }
    \label{fig:mosaic_A}
\end{figure*}

As shown in Table \ref{tab:lenses}, these 58 lenses were then graded by the experts specified in Section \ref{subs:lens_grading}, where the majority vote was taken as the final classification. In total, we identify 16 Grade A lenses, 16 Grade B lenses and 26 Grade C lenses. Figure \ref{fig:mosaic_A} shows a mosaic of the 16 Grade A lenses that we identify in the ASTRODEEP catalogue. Inset to each image is the source coordinates, the redshift and its measurement method (spectroscopic or photometric). We also identify multiple types of lenses in our sample of identified lenses. The first two examples are lenses caused by a cluster - where the arcs cover wide area often larger than other galaxies within the cluster. However, the remaining 14 lenses are strong galaxy-galaxy lenses that are found in different fields.

The majority of the lenses we identify are lensed by early-type galaxies, as is often expected \citep[][]{2006ApJ...638..703B}. We identify no complete Einstein rings, however, many have the correct geometric alignment to form arcs which almost connect to each other (e.g. Cutout 3, 5, 7 and 12 of Figure \ref{fig:mosaic_A}). The photometric redshift of each lens is taken from the ASTRODEEP catalogue, with a single source photometric redshift being found in \citet[][]{Busarello_2002} as it is a cluster lensing system (Cutout 1). Spectroscopic redshifts also come from this catalogue which are collected from ancillary sources. However, for the majority of sources no spectroscopic information is available.

\begin{figure*}
    \centering
    \includegraphics[width=0.95\linewidth]{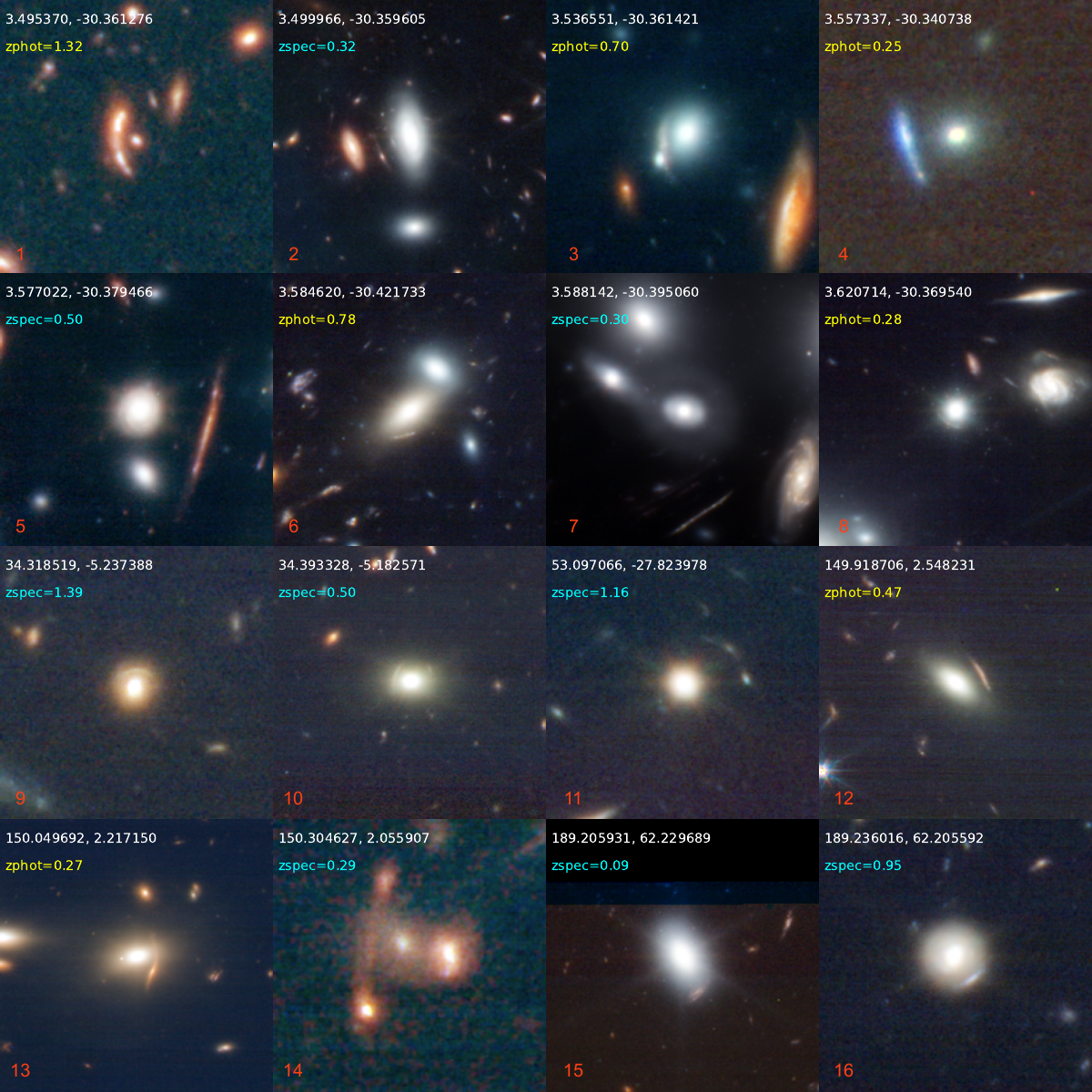}
    \caption{As Figure \ref{fig:mosaic_A}, but showing our grade B lenses. These sources often show approximately the correct geometry but lack either clear colour difference with the lensing source or show a feature that could mean it is a galaxy.}
    \label{fig:mosaic_B}
\end{figure*}

Figure \ref{fig:mosaic_B} shows the lenses identified and classified as Grade B. We are less confident that these sources are gravitational lenses, although they do host the correct geometry to be a lensed arc, they do not show a different colour from their lensing source, their size is not correct or there is some feature in the arc which leads us to believe the arc could actually be a galaxy. For instance, Cutout 14 of Figure \ref{fig:mosaic_B}: the object appears like it could be a gravitational arc but is the same colour as the lensing source and does not bend as expected. However, there are sources such as Cutouts 9, 10 and 13 that potentially even show secondary images, but they lack colour difference or a clear secondary image. The Grade B in Cutout 7 is not the central galaxy of the cutout (which is a ring galaxy), but in fact the small arc feature at the bottom of the cutout. Investigation showed that this is likely an arc in a larger cluster.

\begin{figure*}
    \centering
    \includegraphics[width=0.95\linewidth]{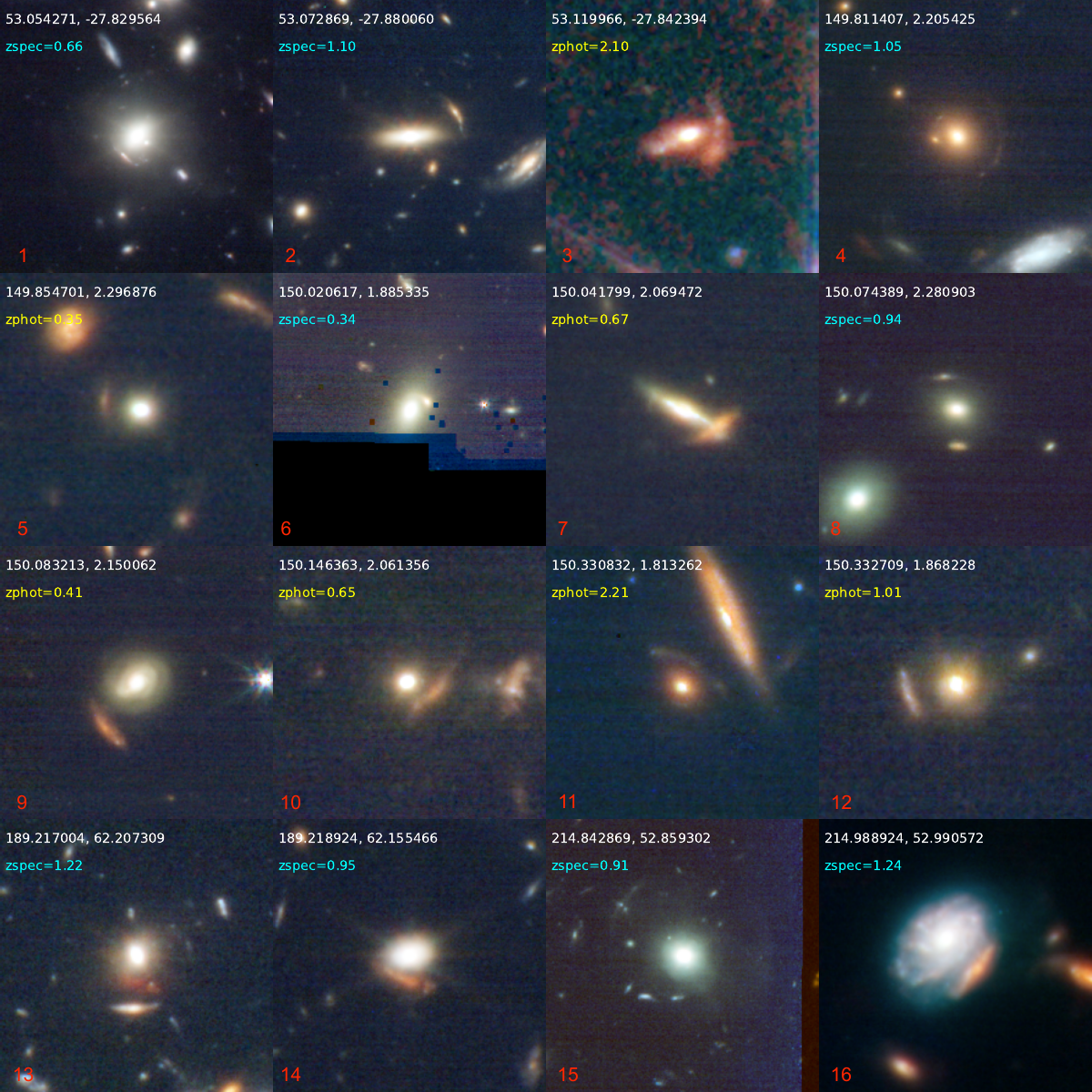}
     \caption[]{As Figure \ref{fig:mosaic_A} but showing our grade C lenses. These images contain sources that are likely to be contamination as they are missing much of the criteria that defines them as gravitational lensing systems.}
     \label{fig:mosaic_C1}
\end{figure*}

\begin{figure}
    \centering
    \includegraphics[width=0.9\linewidth]{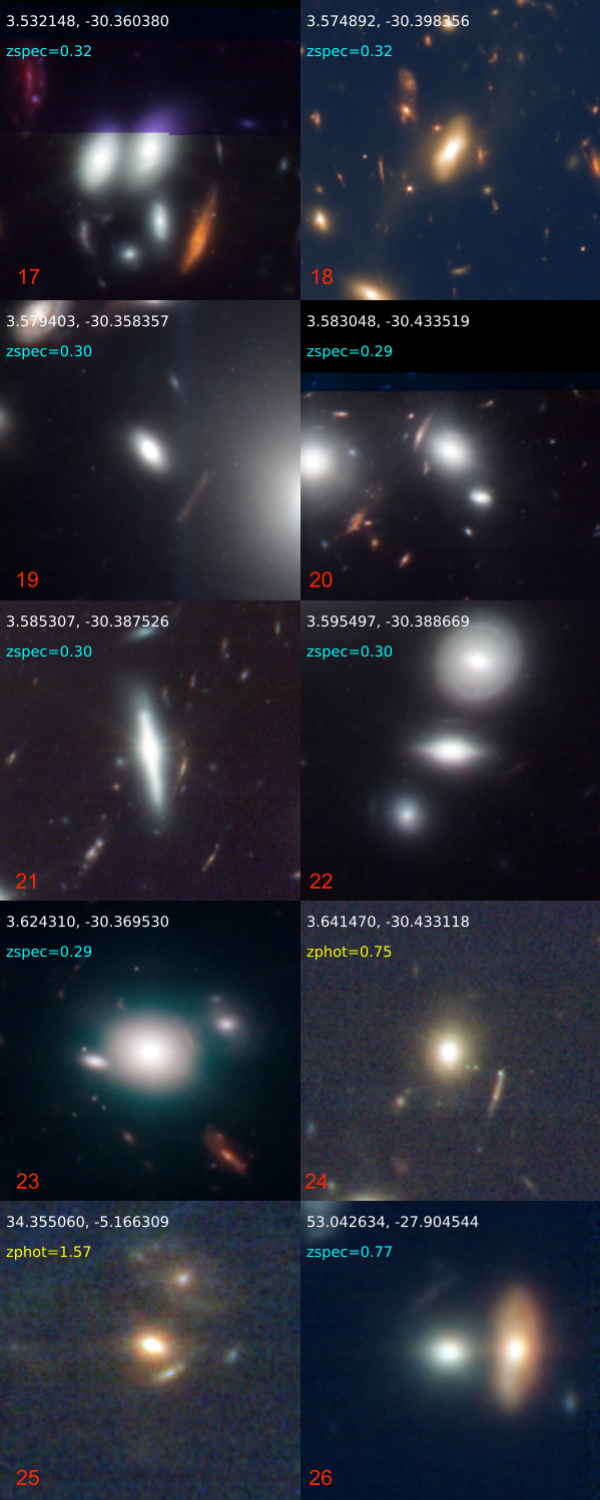}
    \caption{Figure \ref{fig:mosaic_C1} continued}
    \label{fig:mosaic_C2}
\end{figure}

Finally, Figure \ref{fig:mosaic_C1} and \ref{fig:mosaic_C2} shows our grade C gravitational lenses. These are sources where an arc is visible, but this could be a feature not related to gravitational lensing. There are many sources of contamination for gravitational lenses: tidal features from interaction, edge on galaxies with odd morphologies and image artifacts. Therefore, it is important to approach such sources with reasonable scepticism. For instance, many of the examples in this grade are arcs which are the same colour as their lensing source often with a companion, which could be a merger. These examples include cutouts 1, 7 and 10 in Figure \ref{fig:mosaic_C1}. There are also some examples that we can be fairly confident in, however, the image creation makes the arcs particularly faint compared to the lensing source - for example in Cutout 14. We also include objects here where the colour difference between the sources seems clear, but the alignment of the lensing source and the arc is odd. For instance, Cutouts 17 and 19.

Thus, we have identified 58 gravitational lens systems. We also want to identify which of these are new discoveries and which have been identified in previous works. To achieve this, we cross-match against multiple literature catalogues. First, we automatically query the ESASky\footnote{\href{https://sky.esa.int/}{https://sky.esa.int/}} service at each lens coordinates. Using the Python API of ESASky, we are able to return a list of works which have been associated with the source. This list contains each work's title and abstract. We use a search radius of 0.5$^{\prime\prime}$ about each source coordinate and search for keywords in the title and abstract for information if the source has been discussed in the context of gravitational lensing. We use such a precise matching criteria due to the astrometric precision of the \textit{JWST} data being used. If we find that the source has been discussed in the context of lensing, we assign a True value to the referenced column in Table \ref{tab:lenses} and False if the source is discussed in other contexts. To ensure that we are not missing any references for our sources, we also conduct the same process with SIMBAD and the NASA Extragalactic Database (NED).

We find none of our sources have zero references associated with them - which is expected, as they all appear in ASTRODEEP or the many \textit{JWST} surveys which took the observations. However, we find 37 of the 58 systems have no literature associated with claiming them to be gravitational lens systems - 63.7\% of sources. Of the `spectacular' lenses we find - i.e. in Grade A, 9 of the 16 are new discoveries in this work (56.3\%). For Grade B, we find a higher rate of new discoveries of 10 of 16 (62.5\%) and 18 of 26 Grade C lenses (69.2\%). We note, however, that the increasing new discovery at lower grade will also be influenced by an increased level in contamination of our sample.

\begin{figure}
    \centering
    \includegraphics[width=1\linewidth]{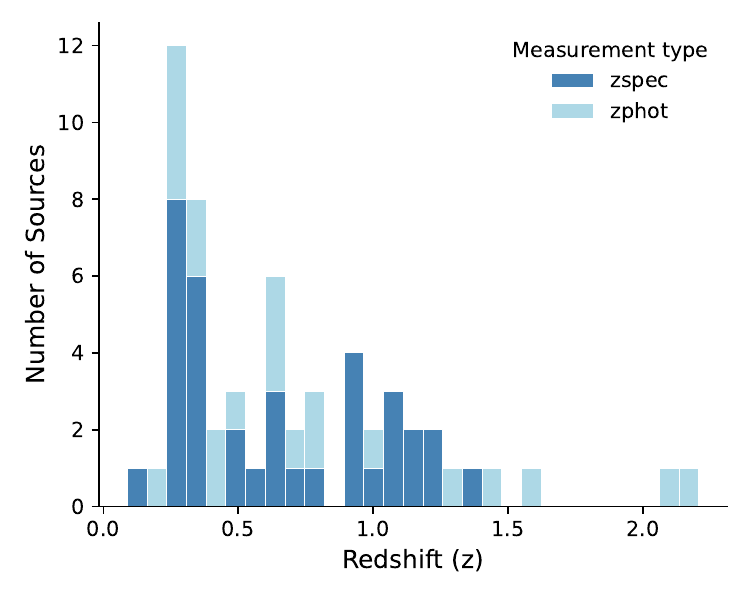}
    \caption{Distribution of combined spectroscopic and photometric redshift measurements for identified and non-identified lenses, limited to the minimum and maximum values of the data. The non-identified sources exhibit a higher median redshift, whereas the identified lenses include the highest $z_\text{lens}$ values.}
    \label{fig:redshift_dist}
\end{figure}

We now investigate the breakdown of the redshift distribution of both the lenses we have recovered from previous works and those we have identified in this work. Figure \ref{fig:redshift_dist} shows the redshift distribution vs number of systems. We break down the redshift measurements into photometric and spectroscopic redshifts. The highest redshift object we find (which was found previously in COWLS) is at $z_{phot}=2.21$. The highest redshift lens we identify which is newly discovered in this work is at $z_{phot}=2.1$. This is a Grade C lens with the cutout being shown in Cutout 3 of Figure \ref{fig:mosaic_C1}. As this is a photometric redshift, this high redshift system must be taken lightly as a spectroscopic redshift measurement may change the measurement. The highest spectroscopic redshift lens we identify is $z_{spec}=1.390$ and is a Grade B lensing system. This is shown in Figure \ref{fig:mosaic_B} in Row 3, first image. While graded as B here by our strict criteria, we can confidently say that this is a lens due to its geometry. However, its colour does match that of the lensing source.

Finally, we investigate the difference in the redshift distributions of the lensing systems we have newly found in this work vs those that were previously known in the literature. The systems identified previously are usually in works using telescopes that do not have the depth or resolution of JWST. Therefore, we would anticipate that we identify lensing systems over a wider redshift range than previous works. Figure \ref{fig:redshift_violin_plot} shows the difference in the redshift distributions as a violin plot. This clearly shows the redshift distribution of the lensing systems found is significantly more broad across redshift compared to previous works (bearing in mind that the previously known sample reflects what earlier searches have detected). 

\section{Discussion}\label{discussion}
\begin{figure}
    \centering
    \includegraphics[width=0.9\columnwidth]{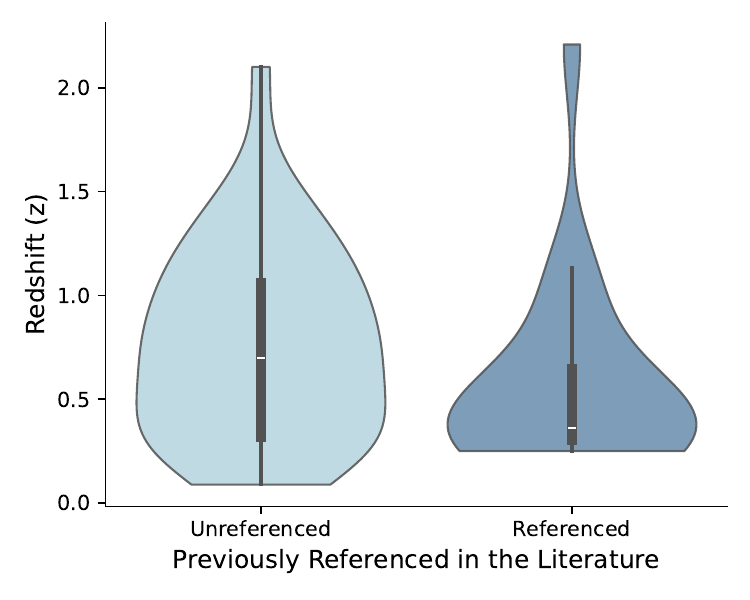}
    \caption{Violin plots showing the redshift distribution of lensing systems discovered in this work (\textit{Right}: Unreferenced) and those that were previously found in the literature (\textit{Left}: Referenced).}
    \label{fig:redshift_violin_plot}
\end{figure}

\noindent This work demonstrates the power of \textit{JWST} to identify gravitational lenses to significantly higher redshifts than previously achieved with other observatories \citep[][]{2006ApJ...638..703B, 2020ApJ...894...78H}. Applying \anomalymatch{} to the ASTRODEEP and COSMOS-Web photometric catalogue across $\sim 0.74\mathrm{deg}^2$ of \textit{JWST} survey area, we identify 58 lenses, of which 63.8\% are newly discovered. As shown in Figure \ref{fig:redshift_violin_plot}, the redshift distribution of these systems spans a significantly wider range than previously identified lenses in the same survey area, stretching from z$\sim 0.089$ to 2.2. This extends the findings of COWLS \citep[][]{nightingale2025, mahler2025}, which demonstrated the potential of \textit{JWST} lens searches in the COSMOS-Web footprint, identifying lenses as high as $z_{\text{phot}} = 2.208$. 

Taken together, these results highlight that deeper, smaller-area \textit{JWST} surveys are a powerful complement to wide-field searches such as those conducted with SDSS \citep[][]{2006ApJ...638..703B}, DESI \citep[][]{2019MNRAS.484.5330J}, and those forecast for LSST \citep[][]{2015ApJ...811...20C} and \textit{Euclid} \citep[][]{2025NatAs...9.1116L, 2025arXiv250315326E} - not only extending sky coverage but pushing lens identification to higher redshifts.

A key result of this work is the demonstration that semi-supervised anomaly detection with \anomalymatch{}, deployed via ESA Datalabs, provides an efficient and scalable pipeline for gravitational lens searches in \textit{JWST} data. The science platform enabled straightforward cross-matching of all ASTRODEEP sources with their Level 3 observations and cutout generation at scale. Looking ahead, it would be feasible to extend this pipeline to the entire \textit{JWST} science archive using the same platform. This would introduce additional complexity - in particular, consistent source extraction and size estimation across heterogeneous observations taken for different purposes, with varying exposure times and background estimations - which is challenging but tractable. In this way, science platforms such as ESA Datalabs open the prospect of scanning historic archival data for gravitational lenses that have previously been missed, significantly amplifying the scientific return of existing observations. 

The graded catalogue produced here - spanning grades A through C reflecting author confidence - provides a useful resource for future ML-based lens searches in JWST. However, several caveats should be noted. Our training set consisted solely of the `spectacular' lenses from COWLS, chosen to minimise contamination from morphologically similar non-lenses such as spiral arms, galaxy mergers, galaxies undergoing ram pressure stripping, and image artefacts. While this choice was deliberate - contamination in the labelled set is particularly damaging for semi-supervised methods that heavily augment it - it introduces a bias towards the clearest, most unambiguous lensing geometries. Subtler systems may have been missed. The grade classifications themselves were determined by a small panel of experts via majority vote and should therefore be treated as an indication of classifier confidence rather than a definitive classification. Those wishing to use this catalogue as a training set for future searches should select sources appropriate to the specific systems they wish to identify, rather than adopting the full catalogue uncritically. 

\section{Conclusions}\label{conclusion}
\noindent In this work, we apply the novel semi-supervised machine learning tool \anomalymatch{} to search \textit{JWST} archival data for strong gravitational lenses. Using the ASTRODEEP and COSMOS-Web photometric catalogues, which span the JADES, CEERS, NGDEEP, PRIMER, Abell 2744 and COSMOS-Web fields. We generate 614,015 three-channel colour cutouts via MTF normalisation of the $F115W$, $F150W$, $F277W$, and $F444W$ NIRCam bands, leveraging the ESA Datalabs science platform for efficient crossmatching and data access. Training \anomalymatch{} on just 11 `spectacularly' confirmed lenses from COWLS \citep{nightingale2025}, we identify 58 gravitational lens candidates spanning a redshift range up to $z_\text{phot} = 2.1$ - demonstrating that label-efficient semi-supervised methods can recover lenses at higher redshifts than previously achieved with other observatories.

Applying our trained \anomalymatch{} model across the ASTRODEEP and COSMOS-Web compiled surveys reveals 58 gravitational lens systems. Upon crossmatching our sources with the literature, we find 37 (63.8\%) are new and have not been discussed as gravitational lens systems. We present each of these lenses in different mosaics ordered by their grading. This grade, between A and C, is indicative of the quality and confidence we have in the lenses we have found. These were calculated using a majority vote of the authors. We find 16 lenses of Grade A, 16 of Grade B and 26 Grade C lenses. While we are confident in the Grade A lenses identified, we caution that the Grade B and Grade C classifications may contain contamination when used for further study.

We investigate the redshift distribution of the new lenses we have found and identify them to a higher redshift than the previously found works. The highest redshift lens we identify in this search is $z_{\text{phot}} = 2.1$, comparable to the visual search conducted by COWLS \citep{nightingale2025}.

Thus, \anomalymatch{} shows that it fills a unique niche for the purpose of gravitational lens detection: a label-efficient detection methodology even under severe class imbalance. Due to its semi-supervised methodology combined with active learning, we are able to train from very few initial labels and begin identifying far more lenses in the \textit{JWST} archives. This could not only be applied to gravitational lens searches, but to other rare objects of interest where labels are few.  

Finally, this work shows not only the power of \textit{JWST} to identify lenses to a deeper redshift distribution than previous observatories, but also the power of science platforms such as ESA Datalabs. Using this, and its direct access to the data, we are able to scan terabytes of archival data quickly and efficiently to search for gravitational lenses. This platform combined with novel semi-supervised methods shows a powerful and scalable capability to identify rare objects of interest across the archives. While we have only applied our search to those surveys contained in ASTRODEEP, ESA Datalabs allows us access to all archival data from JWST. Thus, future work will combine our approach with efficient source extraction via ESA Datalabs to search the entire \textit{JWST} archive for high-redshift gravitational lenses.

We release these gravitational lenses in this work, their coordinates and cutout images to the community via both the CDS and Zenodo (\href{https://zenodo.org/uploads/19147582}{10.5281/zenodo.19147582}). This lens sample can not only be a training set for other machine learning algorithms, but can also be the foundation of further work to search through the entire \textit{JWST} archives. 

\begin{acknowledgements}
     JD acknowledges this work was supported by the ESA Archival Research Visitor Program. DOR acknowledges the support of the ESA Research Fellowship in Space Science program. This research was published as part of that fellowship. This work made extensive use of the ESA Datalabs platform that is described in \citet{navarro2024}. This work is based [in part] on observations made with the NASA/ESA/CSA James Webb Space Telescope. The data were obtained from the Mikulski Archive for Space Telescopes at the Space Telescope Science Institute, which is operated by the Association of Universities for Research in Astronomy, Inc., under NASA contract NAS 5-03127 for JWST. The identification of sources in these surveys was from the ASTRODEEP photometric catalogues, fully described in \citet{merlin2024} and COSMOS-Web catalogues fully described in \citet{casey2023}, with the specific imaging of NIRCam described in \citet{2026ApJ...999..200F}. 

     We especially thank those who took the time to aid in the classification of lenses in the Zooniverse project: including all co-authors. Experiments for how best to balance and create the cutouts were conducted during the Galaxy Zoo Collaboration meeting at the International Space Science Institute in Bern, Switzerland. We acknowledge their support of such a meeting the helpful discussion with the Galaxy Zoo team.
     
     To study the objects we found, and aid in their classifications, extensive use of the ESASky platform and SIMBAD database were used. The ESASky platform was developed by the ESAC Science Data Centre (ESDC) team and maintained alongside other ESA science mission's archives at ESA's European Space Astronomy Centre (ESAC, Madrid, Spain). A full description of the platform can be found in \citet{2017PASP..129b8001B} and \citet{2018A&C....24...97G}. This research has also made use of the SIMBAD database, operated at CDS, Strasbourg France. It is fully described in \citet{2000A&AS..143....9W}.

     The semi-supervised method used to identify lenses in this work was the \anomalymatch{} method developed at the European Space Agency. This is available on GitHub at \href{https://github.com/esa/AnomalyMatch}{https://github.com/esa/AnomalyMatch} and described in \citet{gomez2025}.

     This research made use of many open-source Python packages and scientific computing systems. These included \texttt{Matplotlib} \cite{matplotlib}, \texttt{Pandas} \citep{pandas},and \texttt{numpy} \citep{numpy}. This work also extensively used the community-driven Python package \texttt{Astropy} \citep{astropy:2018, astropy:2022}.

     AI tools were used in the development of the code for lens identification and analysis. AI tools were also used in the editing of this manuscript. 
\end{acknowledgements}
\bibliographystyle{aa}
\bibliography{references}

\end{document}